\documentclass[12pt]{article}
\pdfoutput=1
\usepackage{latexsym}
\usepackage{epsfig,amssymb,euscript}
\usepackage{amsmath}
\usepackage{mathrsfs}
\usepackage[margin=30pt,font=small,labelfont=bf]{caption}
\numberwithin{equation}{section}  
\newsavebox{\ns}
\newsavebox{\dbrane}

\def\be{\begin{equation}}
\def\ee{\end{equation}}
\def\bea{\begin{eqnarray}}
\def\eea{\end{eqnarray}} 

\renewcommand{\theequation}{\arabic{section}.\arabic{equation}}
\def\theequation{\thesection.\arabic{equation}}

\def\Dslash{\,\,{\raise.15ex\hbox{/}\mkern-12mu D}}
\def\Dbarslash{\,\,{\raise.15ex\hbox{/}\mkern-12mu {\bar D}}}
\def\delslash{\,\,{\raise.15ex\hbox{/}\mkern-9mu \partial}}
\def\delbarslash{\,\,{\raise.15ex\hbox{/}\mkern-9mu {\bar\partial}}}
\def\pslash{\,\,{\raise.15ex\hbox{/}\mkern-9mu p}}
\def\calDslash{\,\,{\raise.15ex\hbox{/}\mkern-12mu {\cal D}}}

\newcommand{\nn}{\nonumber \\}
\newcommand{\reef}[1]{(\ref{#1})}

\begin{document}

\makeatletter
\renewcommand{\theequation}{\thesection.\arabic{equation}}
\@addtoreset{equation}{section}
\makeatother

\baselineskip 18pt

\begin{titlepage}

\vfill

\begin{flushright}
Imperial/TP/2012/JS/02\\
\end{flushright}

\vfill

\begin{center}
   \baselineskip=16pt
   {\Large\bf  Bosonic Fractionalisation Transitions}
  \vskip 1.5cm
      Alexander Adam$^{1)}$, Benedict Crampton$^{1)}$, Julian Sonner$^{1),2)}$,  Benjamin Withers$^{3)}$\\
   \vskip .6cm
        \begin{small}
      {}$^{1)}$\textit{Theoretical Physics Group, Blackett Laboratory, \\
        Imperial College, Prince Consort Rd, London SW7 2AZ, U.K.}
        \vskip2em
        {}$^{2)}$\textit{DAMTP, University of Cambridge, \\  C.M.S. Wilberforce Road, Cambridge, CB3 0WA, U.K.}
         \vskip2em
        {}$^{3)}$\textit{Centre for Particle Theory \& Department of Mathematical Sciences,\\ Science Laboratories, South Road, Durham DH1 3LE, U.K.}
        \end{small}\\*[.6cm]
   \end{center}

\vfill

\begin{center}
\textbf{Abstract}
\end{center}

\begin{quote}
At finite density, charge in holographic systems can be sourced either by explicit matter sources in the bulk or by bulk horizons. In this paper we find bosonic solutions of both types, breaking a global $U(1)$ symmetry in the former case and leaving it unbroken in the latter. Using a minimal bottom-up model we exhibit phase transitions between the two cases, under the influence of a relevant operator in the dual field theory. We also embed solutions and transitions of this type in M-theory, where, holding the theory at constant chemical potential, the cohesive phase is connected to a neutral phase of Schr\"odinger type via a $z=2$ QCP .
\end{quote}

\vfill

\end{titlepage}
\setcounter{equation}{0}


\section{Introduction}
Given a  holographic system at finite density with respect to some U(1) symmetry, we would like to answer basic questions relating to the physical nature of its groundstate such as {\it `is the symmetry broken or unbroken?'} and in the latter case {\it `are there Fermi surfaces present?'} and {\it `what are the properties of such Fermi surfaces?'}. Lately such questions have attracted a great deal of interest leading to general statements along the lines of a holographic Luttinger relation and its bosonic analogue \cite{Hartnoll:2011pp,Huijse:2011hp,Iqbal:2011bf}. The Luttinger count relates the sum of the volumes of all Fermi surfaces present to the total charge, while its bosonic analogue relates the total charge to the `Magnus force' felt by a test vortex in the dual field theory. Such studies were guided by the realisation that the finite charge density of a field theory - as encoded in the dual geometry via a finite electric flux - can be sourced either by explicit charged matter, such as fermions or charged bosonic condensates, or by a charged horizon. The latter situation corresponds to having some of the charge density of the field theory tied up in gauge-variant operators, and correspondingly not contributing to a naively defined Luttinger count or its bosonic analogue. The charge contributing to this deficit is identified with so-called `fractionalised' charge in the field theory, that is charge carried by deconfined `quark' degrees of freedom which are charged under the gauge symmetry. This point of view has recently been bolstered by the identification of the corresponding $2k_F$ singularities associated with the `hidden' fermionic degrees of freedom by Faulkner and Iqbal \cite{Faulkner:2012gt}.

There can be interesting phase transitions between situations in which flux is sourced by a horizon to ones where it originates from explicit matter sources in the bulk \cite{Hartnoll:2011pp,Puletti:2010de,D'Hoker:2012ej}. An early example of such a transition was observed in the M-theory superconductor \cite{Gauntlett:2009dn,Gauntlett:2009bh}, where it was demonstrated that the zero-temperature limit of the superconducting black hole was a charged domain-wall geometry without horizon due to the interplay of the potential and effective gauge coupling as defined by a neutral scalar field.

The ground states of the Abelian Higgs model with a W-shaped potential were analysed in \cite{Gubser:2008pf,Gubser:2009cg}, where it was observed that the nature of the ground state can change qualitatively as a function of the gauge coupling. In \cite{Gauntlett:2009dn,Gauntlett:2009bh} it was also pointed out that the disappearance of the horizon as $T\rightarrow 0$ can be related to Nernst's Law since the charged domain wall has zero entropy, whereas the naive extrapolation of the finite-temperature geometry to an extremal black hole would carry entropy. It is thus perhaps expected that the charged domain wall is thermodynamically preferred over the finite-entropy black hole solutions in precisely that region of the phase diagram where both solutions coexist and thus the `entropic singularity' of the M-theory superconductor is cloaked by a superconducting dome.

In this work we explore bosonic analogues of the fractionalisation transition of \cite{Hartnoll:2011pp}, corresponding to transitions from phases with broken U(1) symmetry to phases with intact U(1) symmetry and all charge associated with a horizon. The former constitutes the analogue of the `mesonic phase\footnote{Apart from this one instance we adopt the term `cohesive' for any non-fractionalised phase \cite{Hartnoll:2012wm}, in order to avoid potential confusions with the more conventional meaning of the term `mesonic'.}' of \cite{Hartnoll:2011pp} and the latter corresponds to a fully fractionalised phase. There may also be partially fractionalised phases, in which some of the charge is carried by the horizon and some by the scalar matter. Because of the presence of non-singlet matter, the partially fractionalised phase always breaks the $U(1)$ symmetry, so that the transition between a cohesive and a partially fractionalised phase must occur within the superfluid state. Hence we refer to the two kinds of ordered phases in this paper as the `superfluid cohesive phase' and the `superfluid fractionalised phase'. The transition between the cohesive and fractionalised phase has striking implications for the full phase diagram (see Fig. 3),  yet at this point no field-theoretical order parameter is known for this transition. 
We study these transitions in a variety of models, both in a `bottom-up' and a `top-down' context, with a range of different behaviours, and we describe the conditions under which fractionalisation transition can occur. Among other things we describe a supergravity-embedded superfluid cohesive to neutral phase transition, which has certain aspects that are reminiscent of the standard superfluid to insulator transition in the Bose-Hubbard model at constant chemical potential.

It is worth commenting that the theories considered in this paper have a structural similarity to the Thomas-Fermi approach used in the electron-star literature \cite{Hartnoll:2010gu} with the charged scalar field taking on the r\^{o}le of the fluid of bulk fermions. It should not come as a surprise then that some of the IR geometries encountered in those studies will also feature in the bosonic theories here.

An important r\^{o}le is played in the present work by IR geometries with certain scaling or hyperscaling properties. Such solutions have previously found applications in applied holography, for example in describing QCD like theories \cite{Gursoy:2007cb,Gursoy:2007er,DeWolfe:2010he}, and more recently in describing quantum-critical matter at finite density \cite{Taylor:2008tg,Goldstein:2009cv,Charmousis:2010zz,Shaghoulian:2011aa,Ogawa:2011bz,Huijse:2011ef, Gouteraux:2011ce,Dong:2012se}.

The paper is organised as follows. Immediately following this paragraph we briefly summarise our results. In section $2$ we introduce the class of Einstein-dilaton-charged scalar models used throughout the paper and develop their thermodynamics and holographic renormalisation. Section $3$ gives a detailed discussion of certain bottom-up models which illustrate the general concept of bosonic fractionalisation and motivate the more subtle M-theory case to which we turn in section $4$. We finish with a discussion and a short appendix.

\subsection{Summary of results}
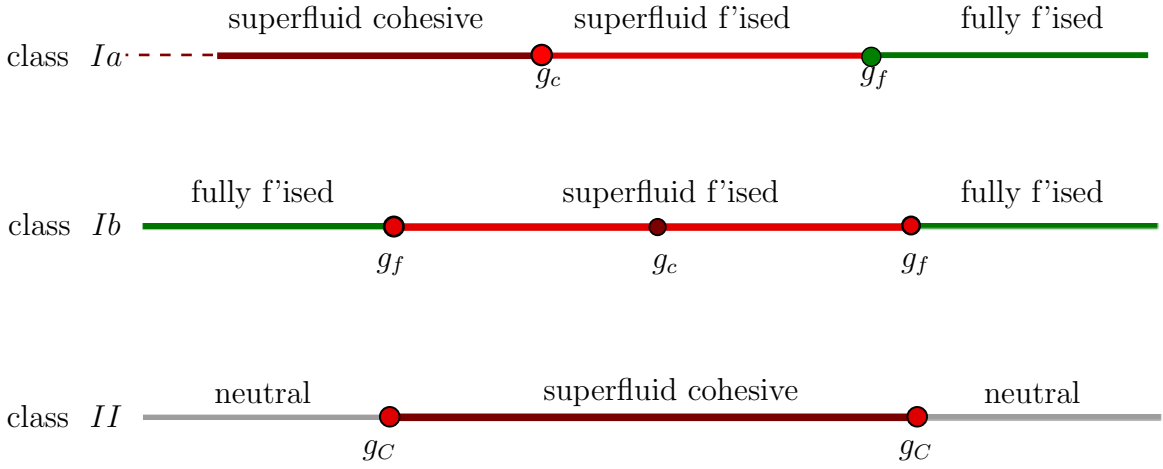
\begin{figure}[h!]
\begin{center}
\includegraphics[width=0.89\textwidth]{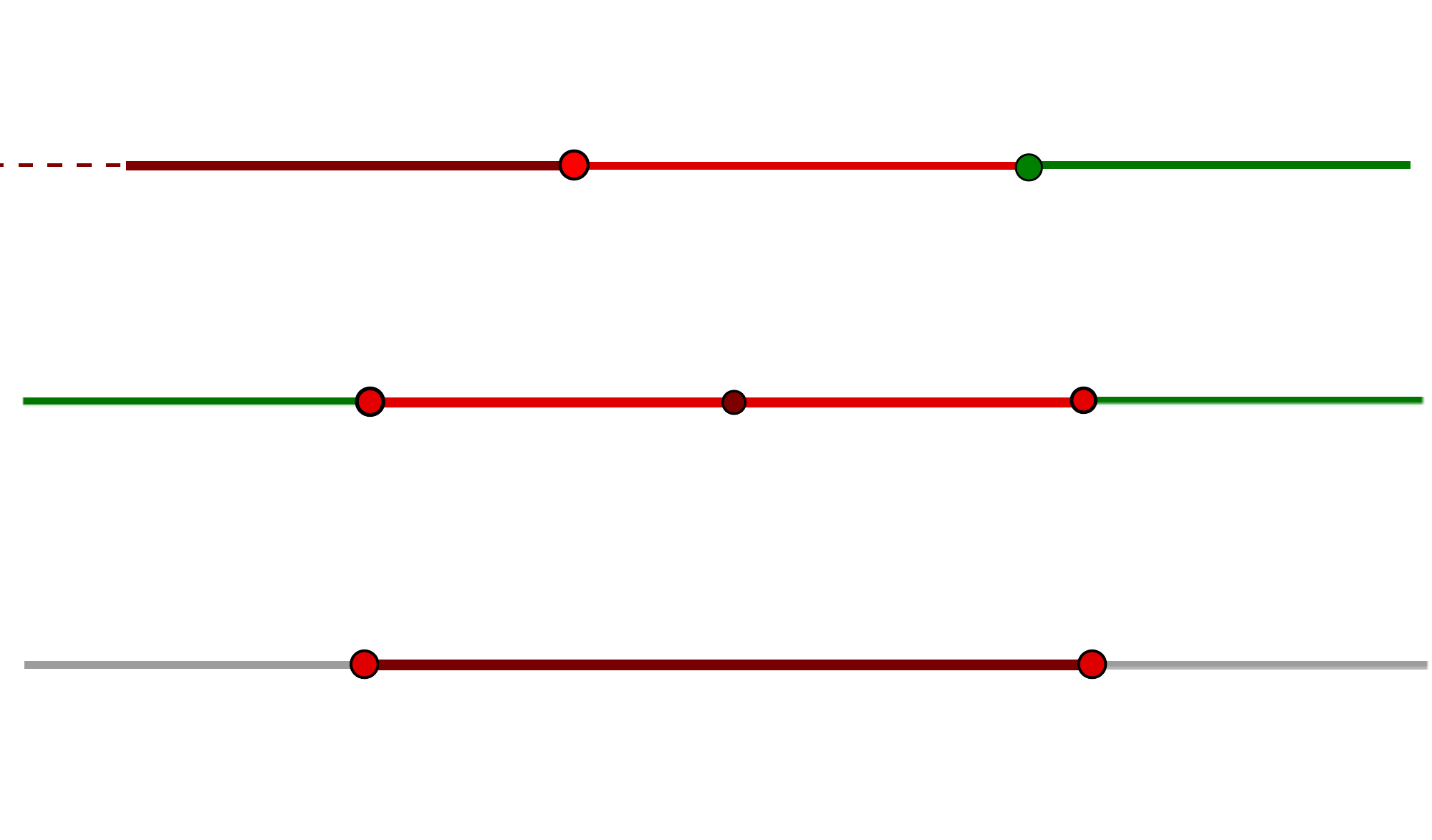}
\setlength{\unitlength}{0.1\columnwidth}
\begin{picture}(0.1,0.25)(0,0)
\put(-9.6,4.2){\makebox(0,0){{\rm class}\,\, $Ia$}}
\put(-9.6,2.75){\makebox(0,0){{\rm class}\,\, $Ib$}}
\put(-9.6,1.1){\makebox(0,0){{\rm class}\,\, $II$}}
\put(-7.1,4.5){\makebox(0,0){\rm superfluid cohesive}}
\put(-5.45,4.0){\makebox(0,0){$g_c$}}
\put(-2.65,4.0){\makebox(0,0){$g_f$}}
\put(-4.3,4.5){\makebox(0,0){\rm superfluid f'ised}}
\put(-1.3,4.5){\makebox(0,0){\rm fully f'ised}}
\put(-7.9,1.3){\makebox(0,0){\rm neutral}}
\put(-4.4,1.3){\makebox(0,0){\rm superfluid cohesive}}
\put(-1.3,1.3){\makebox(0,0){\rm neutral}}
\put(-7.9,3){\makebox(0,0){\rm fully f'ised}}
\put(-4.4,3){\makebox(0,0){\rm superfluid f'ised}}
\put(-1.3,3){\makebox(0,0){\rm fully f'ised}}
\put(-6.9,0.8){\makebox(0,0){$g_{C}$}}
\put(-2.3,0.8){\makebox(0,0){$g_{C}$}}
\put(-6.8,2.4){\makebox(0,0){$g_{f}$}}
\put(-2.3,2.4){\makebox(0,0){$g_{f}$}}
\put(-4.45,2.4){\makebox(0,0){$g_{c}$}}
\end{picture}
\caption{Zero-temperature phase structure of the the models considered in this paper. Shades of red indicate broken $U(1)$ symmetry, with dark red symbolising the cohesive states and bright red the superfluid fractionalised ones. The $U(1)$ symmetry is unbroken in the green fully fractionalised and grey neutral phases. The class $Ib$ example is even in $\Phi$, while the class $Ia$ example has no particular discrete symmetry associated with the dilaton. Note that the M-theory embedding in section \ref{sec:m-theory} is in class $II$.
\label{fig:summary}}
\end{center}
\end{figure}

The salient features of the analysis boil down to the IR structure of the zero-temperature states. We encounter three cases, as illustrated in Fig. \ref{fig:summary}, depending on the structure of the effective gauge coupling, which is controlled by the neutral scalar field. If the gauge coupling vanishes in the IR, we find a fully fractionalised phase, or a superfluid fractionalised phase. If it remains finite, we find a cohesive phase. The $U(1)$ symmetry is broken in the latter two phases by explicit charged matter outside the horizon. In addition to the fully fractionalised phase, there is one further unbroken phase, where the neutral groundstate does not depend on the chemical potential. This `incompressible' phase together with the transition into it from the cohesive phase via a $z=2$ critical point is reminiscent of the superfluid to insulator transition in the Bose-Hubbard model.

\section{General Features}
In this section we introduce the theories leading to quantum phase transitions in terms of a framework that is general enough to cover both the bottom-up constructions of section \ref{sec:bottom-up} and the M-theory compactifications of section \ref{sec:m-theory}. Our general strategy follows a two-pronged approach. Firstly, at zero temperature we will carry out an analysis of possible IR geometries and the deformations which allow them to be connected to the asymptotic $AdS$ configuration universal to all solutions considered in this paper. We are interested both in IR geometries associated with domain wall configurations and IR geometries associated with (extremal) black-hole horizons. This allows us to classify the possible ground states of the holographic theories in this work. Secondly we will extrapolate the known finite-temperature solutions and their dilatonic deformations to very low temperatures. These can lead to a state with broken $U(1)$ symmetry via a condensation instability at some critical temperature $T_c$ or to a state in which this symmetry is intact. In each case we are able to connect them to the appropriate $T=0$ geometry. In this way we can also study other interesting low-temperature signatures, such as the dynamical critical exponent $z$ and the hyperscaling violation exponent $\theta$, in terms of which the low-temperature entropy scales like
\be
\label{eq:entropy scaling}
S \sim T^{\frac{d-\theta}{z}}\,.
\ee
We are able to identify $z$ and $\theta$ from an appropriate (hyper-)scaling IR geometry, matching perfectly to numerical fits of our low-temperature analysis.
\subsection{Action and Field Content}
The class of theories we work with in this paper is described by an action of the form
\bea
\nn\label{eq:GenLag}
S = \int d^{d+1} x \sqrt{-g} \Bigl[R - \tfrac{1}{4}Z_F(\Phi) F_{MN}F^{MN} -  \tfrac{1}{2}(\nabla\Phi)^2  - Z_S(S)|DS|^2 - V(\Phi,|S|)\Bigr] + S_{\rm CS}\,.\nn
\eea
We denote the $d+1$ bulk directions with upper-case Latin $M,N,\ldots$, whilst field-theory coordinates will be denoted with lower-case Greek $\mu,\nu,\ldots$. The field $\Phi$ is a neutral scalar field, which we will refer to as a `dilaton', whereas $S$ is a charged scalar field with covariant derivative $D S = (\nabla  - iq A )S$. We have included $S_{\rm CS}$, which stands for a possible Chern-Simons or axion-like contribution often found in consistent supergravity reductions. Whilst the  solutions we construct do not depend on $S_{\rm CS}$, we note that in certain cases the phase structure may be altered by its presence, for example through the spontaneous formation of stripes.

We  allow the couplings of the gauge field, as well as the charged scalar kinetic terms, to depend on $\Phi$ and $S$ respectively. The interesting transitions observed in the remainder of this paper can be traced back to the interplay of charged and neutral scalar couplings and the potential. Often it is convenient to put the scalar kinetic terms into canonical form. A little thought shows that this is only possible for the magnitude of the complex field. Writing $S$ in a polar decomposition with phase $iq\varphi$ and then using a field redefinition of the magnitude,  we can write the kinetic term of the charged scalar as
\be\label{eq:complexScalarTransform}
Z_S (|S|)|DS|^2 = (\nabla\eta)^2 + q^2 X(\eta)^2 (\nabla\varphi - A)^2\,.
\ee
We should require that $X(\eta)$ has a small-$\eta$ expansion $X(\eta) \sim \eta + {\cal O}(\eta^2)$.  The hyperscaling geometries are generated by a divergent IR dilaton $\Phi\longrightarrow\pm\infty$, and so it is important to distinguish theories where such a divergence can happen from those were it cannot:
\bea\label{eq:Zcases}
{\rm class}\,\, I:\qquad Z_F(\Phi) &\xrightarrow{\quad{\rm IR}\quad}& \infty \,, \qquad \nn
{\rm class}\,\, II:\qquad Z_F(\Phi) &\xrightarrow{\quad{\rm IR}\quad}&  ({\rm always}\,\,{\rm stays}\,\,{\rm bounded})\,.
\eea
Furthermore $Z_F(\Phi)$ may or may not have even symmetry under $\Phi \rightarrow -\Phi$ as the IR limit is approached, again with important implications for the phase structure at zero temperature. In class \textit{I} we are particularly interested in $Z_F(\Phi)$ without this symmetry, so that \textit{e.g.} if $Z_F(\Phi)$ diverges as $\Phi\longrightarrow+\infty$ then it remains bounded when $\Phi$ diverges with the opposite sign. In this manner the dilaton can naturally drive a fractionalisation-type phase transition.

Finally we require that the potential has an expansion
\be
\ell^2 V(\eta,\Phi) = -6- \Phi^2 - 2\eta^2 + \cdots\,,
\ee
ensuring that there exists an $AdS_4$ vacuum solution with associated $AdS$ length $\ell^2$, and that the two scalars\footnote{The seemingly different mass terms are caused by the different normalisations of the respective kinetic terms. We are using a convention that appears to be standard in the literature.} $\Phi$ and $S$ are dual to $\Delta=2$ operators. For the solutions in this paper we can take the scalar $S$ identically real, so that without loss of generality $\varphi \equiv 0$.

Throughout this paper we use the metric ansatz
\be\label{eq:GeneralMetricChoice}
ds^2 = - f(r) e^{-\beta(r)} dt^2 + \frac{dr^2}{f(r)} + \frac{r^2}{\ell^2} \left( dx^2 + dy^2 \right)\,.
\ee
\subsection{Holographic renormalisation and asymptotic charges}\label{sec:UVcharges}
We are interested in solutions that asymptotically approach the $AdS_4$ fixed point in the UV. Such solutions admit a large $r$ expansion
\bea
f &=& \frac{r^2}{\ell^2} + \frac{1}{2}\left(\eta_{1}^2 + \tfrac{1}{2}\Phi_{1}^2  \right) + \frac{\ell G_{1}}{r} + \cdots\,,\nn
\beta &=& \beta_a + \frac{\ell^2 \left( \eta_{1}^2 + \tfrac{1}{2}\Phi_{1}^2 \right)}{2 r^2} + \cdots\,,\nn
\Phi &=& \frac{\ell \Phi_{1}}{r} + \frac{\ell^2 \Phi_{2}}{r^2} + \cdots\,,\nn
\eta &=& \frac{\ell \eta_{1}}{r} + \frac{\ell^2 \eta_{2}}{r^2} + \cdots\,,\nn
A_t &=& \ell e^{-\beta_a/2} \left(\mu - \frac{Q}{r} + \cdots\right)\,.
\eea
By evaluating the asymptotic stress tensor (which involves introducing counterterms to render the renormalised expressions finite) we can deduce the exact relationship of the integration constants introduced here and the physical charges of the dual field theory.
This is achieved by adding the specific counterterm action
\be
S_{\rm ct} = 2 \int_{\partial \Sigma} \sqrt{-\gamma} \left( {\cal K} - \frac{2}{\ell} \right) - \int_{\partial \Sigma} \sqrt{-\gamma}\frac{1}{\ell} \left(|S|^2 + \tfrac{1}{2} \Phi^2  \right)\,,
\ee
where we have chosen the counterterms for the scalar fields such that the fixed $\eta_{1}, \Phi_{1}$ ensemble has a well defined variational principle. ${\cal K} = \gamma^{\mu\nu}{\cal K}_{\mu\nu}$ is the trace of the extrinsic curvature of a constant $r$ hypersurface with unit normal $n^r = f(r)^{\frac{1}{2}}$. We find that the total action
\be
S_{\rm ren} = S+S_{\rm ct}
\ee
is finite, as is the renormalised stress tensor
\be
T_{\mu\nu} = 2 \left[ {\cal K}_{\mu\nu} - {\cal K}\gamma_{\mu\nu} - \frac{2}{\ell}\gamma_{\mu\nu} \right] - \frac{1}{\ell}\left(|S|^2  + \tfrac{1}{2} \Phi^2 \right) \gamma_{\mu\nu}\,.
\ee
From this we can relate the expansion coefficient $G_{1}$ to the energy density\footnote{The quantity $e^{-\beta_a}$ is the boundary speed of light (squared), and so the conversion between ADM mass and energy, and correspondingly energy density, involves a factor of $e^{-\beta_a}$. For this and related reasons it is convenient to set $\beta_a=0$, which we will assume to be the case from now on.}, defined as
\be
\varepsilon =\frac{re^{\beta_a}}{\ell}T_{00} = -\frac{2}{\ell} \left(  G_{1} - \eta_{1} \eta_{2} - \tfrac{1}{2}\Phi_{1} \Phi_{2} \right)\,.
\ee

The two scalar fields play very different r\^{o}les in our system. The dilaton $\Phi$ gives rise to an explicit deformation parameter: we switch on a relevant operator by sourcing it in the dual theory, {\it i.e.} by taking a non-zero value of the UV parameter $\Phi_1$, and then study the behaviour of the theory as we vary this source. In contrast, the field $\eta$ is only ever allowed to condense spontaneously, {\it i.e.} we set $\eta_1=0$.

\subsection{Thermodynamic Relations}
By analytically continuing to Euclidean signature 
\be
t = -i\tau\,,\qquad I_E = - i S_{\rm ren}\,,
\ee
we define the grand canonical potential
\be
I_E = \beta \Omega(\mu,T) = \beta {\rm vol}_2 \omega(\mu,T)\,.
\ee
Evaluated on-shell, the action can be written as a total derivative
\be
{\cal S}_{\rm OS} =i\beta {\rm vol}_2 \int_{r_+}^\infty\frac{d}{dr} \left[\frac{2}{r}\sqrt{-g}g^{rr}  \right]dr\,,
\ee
which vanishes at the horizon, and thus depends only on the asymptotic charges.
Using techniques of \cite{Sonner:2010yx,Bhattacharya:2011tra} one can show that this is the negative of the pressure
\be
\omega(\mu,T)=-P\,.
\ee
We can exploit the symmetries of the background metric to derive a Smarr-Gibbs-Duhem relation. This follows from the vanishing of a certain total derivative\footnote{The on-shell vanishing of the total derivative explains why the authors of \cite{Gauntlett:2009bh,Sonner:2010yx} were able to write the on-shell action as two different boundary terms.} on shell, as we now show. By using the Killing equation in addition to the Ricci identity for the two Killing vectors $\partial_t$ and $\partial_y$ we can derive the relation
\be\label{eq:TotDerivI}
\sqrt{-g} \left( R^t_t - R^y_y \right) =- \partial_r \left[ \frac{e^{-\beta/2}r^2}{2\ell^2} \left(  f\beta' - f' + \frac{2f}{r} \right) \right]\,.
\ee
It is then a simple matter of using the trace-removed Einstein equations to find that on-shell
\be
\sqrt{-g} \left( R^t_t - R^y_y \right)=-\frac{1}{2}\partial_r \left[\sqrt{-g}Z(\phi)A_t F^{tr}  \right]\,,
\ee
so that, upon integrating from the horizon to the UV boundary, we obtain the relation
\be
 \frac{e^{-\beta/2}r^2}{\ell^2} \left(  f\beta' - f' + \frac{2f}{r}\right) \Biggr|_{r_+}^\infty  = \sqrt{-g}Z(\phi)A_t F^{tr} \Bigr|^\infty\,.
\ee
Evaluated on our expansions this gives the Gibbs-Duhem relation
\be
\frac{3}{2}\varepsilon = \mu Q + Ts - \ell^{-1} \left(\eta_{1} \eta_{2} + \frac{1}{2}\Phi_{1}\Phi_{2}\right)\,,
\ee
or, written in terms of the pressure
\be
\varepsilon + P = \mu Q + Ts\,.
\ee
Here we have used that the temperature is given by
\be
T=\frac{e^{(\beta_a-\beta_+ )/2}}{4\pi \ell^2}f'(r) \Bigr|_{r=r_+}\,,\, \qquad (\beta_a=0)\,.
\ee
Moreover, the entropy density is given in terms of the horizon radius as
\be
s = \frac{r_+^2}{4G_N}\,,\qquad ({\rm note}\quad 16 \pi G_N = 1)\,.
\ee
The above derivations continue to make sense for solutions of the theory \reef{eq:GenLag} where $\lim_{r\rightarrow r_{\rm IR}} f'(r)=0$ as $r_{\rm IR}\rightarrow 0$, \textit{i.e.} solutions of vanishing entropy at zero temperature. We also allow the dilaton field to diverge logarithmically\footnote{Specifically the dilaton blows up logarithmically as a function of radial proper distance.} in this limit, as long as it does not interfere with the conditions just mentioned.

\section{Class \textit{I}: A Bottom-Up Model}
\label{sec:bottom-up}

When $Z$ is not symmetric in $\Phi$ there is the possibility of two qualitatively different behaviours assosciated with a dilaton that diverges logarithmically to either plus or minus infinity. In the following models, we will see that in the first case the effective gauge coupling $e^2\sim Z^{-1}$ will go to zero in the IR and in the second it will diverge. There is a continuous transition between a superfluid fractionalised phase and a cohesive phase, via a quantum critical point with a finite dilaton. Both phases will be of hyperscaling form. This phase transition will be driven by the asymptotic value of the dilaton. We will also see an additional phase transition between a partially and a fully fractionalised phase, assosciated with the edge of the superconducting dome.

We will consider the class of models
\be
	Z_F(\Phi) = Z_0^2 e^{a\Phi/\sqrt{3}}\,,\qquad 
	\ell^2 V(\Phi,\eta) = -V_0^2 \cosh\left(b\Phi/\sqrt{3}  \right) - 2 \eta^2 + g_\eta^2 \eta^4\,,\label{eq:ZV}
\ee
where $a,b >0$.
The $\eta^4$ term allows for interpolating soliton solutions between the $AdS_4$ maximum at $\eta=0$ and the $AdS_4$ minimum at $\eta^2=g_\eta^{-2}$. We will study IR geometries for general values of $a$ and $b$ (see also  \cite{Hartnoll:2011pp,Hartnoll:2012wm}). All our numerical results will be for the particular case $a=b=g_\eta^2=1$, $V_0^2=6$ and $Z_0^2=1$.


\subsection{Zero Temperature}
\label{sec:T=0}

We now study the phases of our system at zero temperature. We construct the IR geometries by making a scaling ansatz for the fields, allowing for a possible logarithmic divergence of the dilaton. With this divergence the potential is schematically $V\sim r^{\delta}+r^{-\delta}$, for some $\delta$, and so these scaling solutions are not exact: they are the leading order parts of series solutions, with subleading corrections in increasing powers of $r$. For each of our three cases we also identify the possible irrelevant deformations.

The divergent dilaton endows our leading order solutions with hyperscaling symmetry \cite{Gouteraux:2011ce,Ogawa:2011bz,Huijse:2011ef}, which is a generalisation of the familiar Lifshitz scaling:
\bea
	\mathbf{x}&\rightarrow& \lambda \mathbf{x}\,,\nn
	t &\rightarrow& \lambda^z t\,,\nn
	r&\rightarrow& \lambda^{(\theta-2)/2}r\,,
\eea
under which $ds\rightarrow\lambda^{\theta/2}ds$. Here $z$ is the usual dynamical critical exponent, and $\theta$ is the hyperscaling violation parameter.

\subsubsection{Fractionalised IR Solutions}
In this case some or all of the flux is sourced by the horizon. Starting with the fully fractionalised case, we switch off the condensate. We find the following scaling solutions at leading order:
\bea
	f(r) e^{-\beta(r)} &=& r^{\frac{2(12 + a^2 - b^2)}{(a+b)^2}}\,,\nn
	f(r) &=& f_0 r^{\frac{2(a - b)}{a+b}}\,,\nn
	A_t(r) &=&A_0 r^{\frac{12 + (3a-b)(a+b)}{(a+b)^2}}\,,\nn
	\Phi(r) &=& -\frac{4\sqrt{3}}{a+b}\log r\,,
\eea
with
\bea
	f_0 &=&\frac{(a+b)^4 V_0^2}{4(6 + a (a+b))(12 + (3a-b)(a+b))}\,,\nn
	A_0^2 &=&\frac{4(6-b(a+b) )}{(12 + (3a-b)(a+b))Z_0^2}\,.
\eea
These solutions have dynamical critical exponent $z$ and hyperscaling parameter $\theta$ given by
\be
	z=\frac{12 + (a-3b)(a+b)}{a^2 - b^2}\,,\qquad \theta=\frac{4b}{b-a}\,.
\ee
Note that though these parameters are infinite in our special case $a=b=1$, their ratio is still finite (such a situation is discussed further in \cite{Hartnoll:2012wm}). This ensures that for solutions close to this $T=0$ geometry, the various thermodynamic parameters have finite scaling exponents with temperature.

There are two irrelevant deformations about this solution. The first simply corresponds to a shift in the IR vacuum expectation value of the charged scalar:
\be
	\delta\eta=\eta_0\,.
\ee
Integrating this deformation to the UV gives geometries with explicit bulk charge in addition to horizon charge.
We thus find that the fully fractionalised solutions connect continuously to the superfluid fractionalised phase branch of solutions. The second deformation is assosciated with the dilaton, and has the form
\be
	\delta\Phi=a_\Phi^{IR}\,r^{\nu_\Phi}\,, \qquad
	\delta f=a_f^{IR}\,r^{\nu_f}\,, \qquad
	\delta\beta=a_\beta^{IR}\,r^{\nu_\Phi}\,, \qquad
	\delta A_t=a_A^{IR}\,r^{\nu_A}\,,
\ee
with exponents
\bea
	\nu_f&=&\nu_\Phi + \frac{2(a-b)}{a+b}\,,\nn
	\nu_A&=& \nu_\Phi + \frac{12 + (3a-b)(a+b)}{(a+b)^2}\,,
\eea
where the coefficients $a_{I}$ and the exponent $\nu_\Phi$ are consistently determined by the IR limit of the linearised equations of motion. For example, in the simplified case $a=b$ the theory admits an irrelevant deformation with exponent
\be
	\nu_\Phi = \frac{3+b^2}{6b^4}\left(-3+\sqrt{81-24b^2}\right)\,,
\ee
and coefficients
\be
	\left\{a_f\,,a_\beta\,,a_A\,,a_\Phi\right\}
		=a^{IR}\left\{-f_0\,,2\,,\frac{A_0\nu_\Phi b^2}{2(3-b^2)}\,,-\frac{b}{\sqrt{3}}\right\}\,,
\ee
with an overall free magnitude $a^{IR}$. This is to be compared with the analogous mode in \cite{Hartnoll:2011pp}.

\subsubsection{Critical IR Solution}
The critical IR solution has non-divergent dilaton (in fact $\Phi=0$) and is just the $AdS_4$ minimum  that arises when the condensate sits at the bottom of its potential, $\eta=\pm g_\eta^{-1}$, with vanishing gauge field $A_t(r)=0$. Consequently its $AdS$ length scale is shifted away from the usual value by a $g_\eta$-dependent factor:
\bea
	f(r)e^{-\beta(r)} &=& r^2 \,,\nn
	f(r) &=& \frac{r^2}{R_{IR}^2} \,,\nn
	\eta(r) &=& \pm g_{\eta}^{-1}\,,
\eea
with
\be
	R_{IR}^2 = \frac{6g_\eta^2\ell^2}{1+g_\eta^2V_0^2}\,.
\ee
There are two irrelevant deformations about this solution: one involving the flux and one the charged condensate. They have the form:
\be
	\delta A_t=a_A^{IR} r^{\nu_A}\,,\qquad \delta\eta = a_\eta^{IR} r^{\nu_\eta}\,,
\ee
with exponents
\bea
	\nu_A &=& \frac{1}{2}\left(-1 + \sqrt{1+\frac{48\ell^2q^2}{Z_0^2\left(1+g_\eta^2V_0^2\right)}}\right)\,, \nn
	\nu_\eta &=& \frac{3}{2}\left(-1 + \sqrt{1+\frac{32g_\eta^2}{3\left(1+g_\eta^2V_0^2\right)}}\right)\,.
\eea

\subsubsection{Cohesive IR Solution}

We seek cohesive solutions by demanding that the horizon flux $\lim_{r\to0}\sqrt{-g}Z(\Phi)F^{rt}$ vanishes. At leading order we find the following scaling solutions and their deformations:
\bea
	f(r)e^{-\beta(r)} &=& r^{2} \,,\nn
	f(r) &=& f_0 r^{\frac{2(3-b^2)}{3}} \,,\nn
	A_t(r) &=& 0 \,,\nn
	\Phi(r) &=& \frac{2b}{\sqrt{3}}\log r\,,
\eea
with
\be
	f_0=\frac{3V_0^2}{4(9-b^2)}\,.
\ee
These solutions have dynamical critical exponent $z$ and hyperscaling parameter $\theta$ given by
\be
z=1\, ,\quad \theta = -\frac{2b^2}{3-b^2}\,.
\ee
Just as in the fractionalised case, the vacuum expectation value of the charged scalar is not fixed, $\delta\eta=\eta_0$. There is then a second irrelevant deformation which introduces the required flux. For $a=b$ this is
\be
	\delta A_t=a_A^{IR}r^{\frac{3+b^2}{6}+\nu_A(\eta_0)}\,.
\ee
This exponent depends on the IR value of the charged scalar:
\be
	\nu_A=\frac{3+b^2}{6}\left(-2+\sqrt{1+\frac{72q^2\eta_0^2}{Z_0^2f_0(3+b^2)^2}}\right).
\ee
This is merely the leading order contribution to the series solution from the deformation; there are additional subleading corrections in powers of $2\nu_A$. We thus have the non-trivial consistency condition that this power be positive, which translates into the constraint on the condensate:
\be
	\eta_0^2\,>\,\frac{(3+b^2)^2f_0Z_0^2}{24q^2}\,.
\ee
A similar structure was seen in the zero-temperature superconducting solutions of \cite{Horowitz:2009ij}.


\subsection{Finite Temperature}

In order to complement our zero-temperature picture, and to complete our phase diagram, we would like to study how these solutions extend to finite temperature. Holographically, a non-zero temperature is indicated by the presence of a (non-degenerate) bulk horizon. We thus make the ansatz
\bea
	\label{eq:FiniteTIR}
	f&=& f_+ (r-r_+) + \cdots \,,\nn
	\beta&=& \beta_+ +  \cdots \,,\nn
	A_t &=& A_{t,+} (r-r_+)+ \cdots \,,\nn
	\Phi &=& \Phi_+ + \cdots \,,\nn
	\eta&=& \eta_+ + \cdots\,.
\eea
Temperature masks the singularities of the solutions of section \ref{sec:T=0}, and the dilaton is finite here. Despite this we will see clearly the imprint of the zero temperature IR scaling behaviour at finite temperature. In particular, our results will exhibit the striking influence that the quantum critical point has on the finite temperature physics.

Such solutions exhibit a superconducting instability \cite{Gubser:2008px,Hartnoll:2008vx}, whose critical temperature depends on the value of the dilaton deformation \cite{Gauntlett:2009bh}. This critical temperature can be dialled to zero, forming the edge of a superconducting dome. We have checked that the free energy of the broken phase solution is lower than that of the unbroken phase whenever it exists.

Since we are holding the theory at finite chemical potential, it is natural to consider the phase diagram as a function of $\Phi_1/\mu$.
\begin{figure}[h!]
\begin{center}
\includegraphics[width=0.99\textwidth]{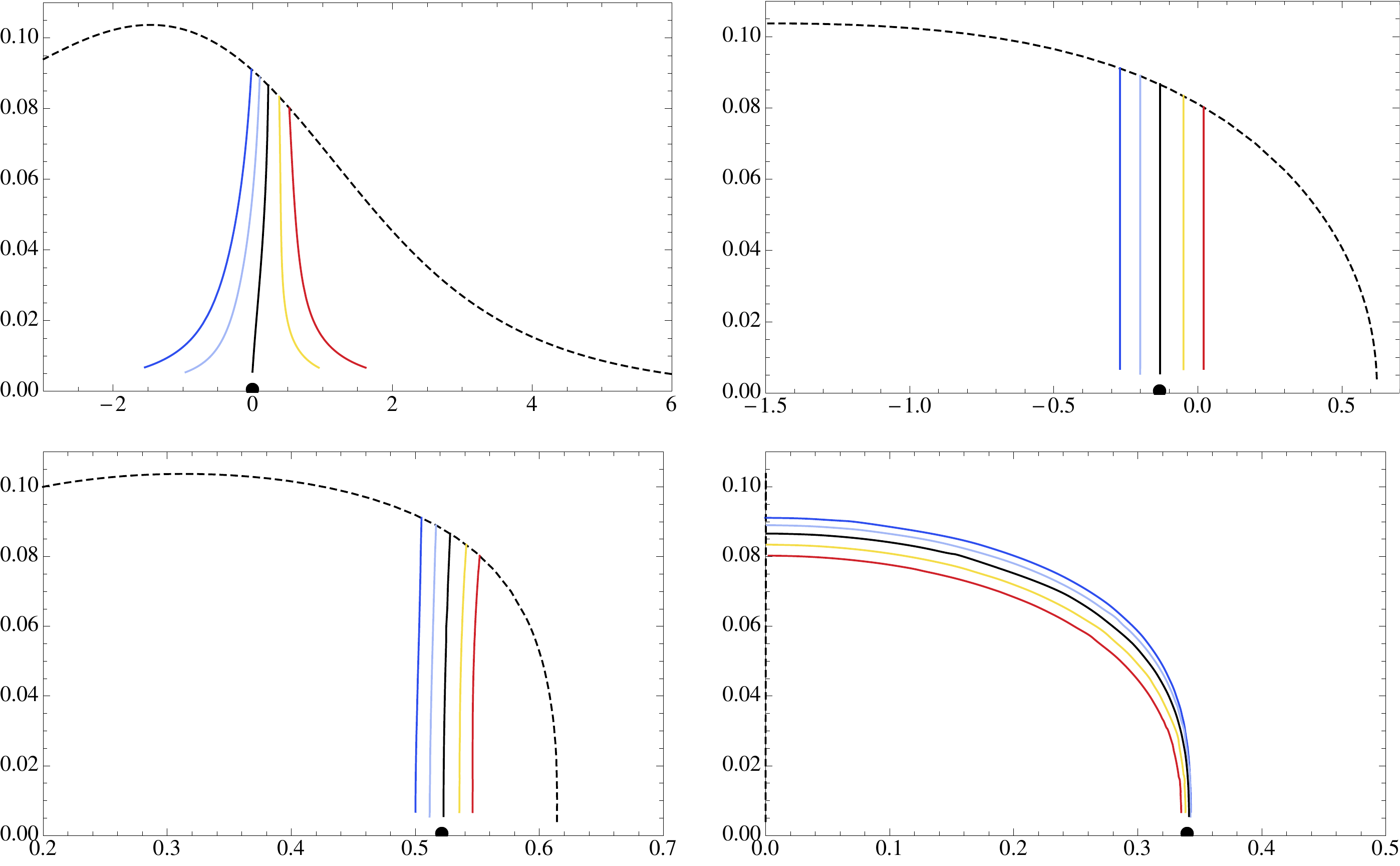}
\setlength{\unitlength}{0.1\columnwidth}
\begin{picture}(0.1,0.25)(0,0)
\put(-5.15,5.8){\makebox(0,0){$\frac{T}{\mu}$}}
\put(-1.3,3.8){\makebox(0,0){$\Phi_{+}$}}
\put(4.2,3.8){\makebox(0,0){$\tfrac{\Phi_{1}}{\mu}$}}
\put(-1.3,0.7){\makebox(0,0){$\tfrac{Q}{\mu^2}$}}
\put(4.2,0.7){\makebox(0,0){$\tfrac{\eta_2}{\mu^2}$}}
\end{picture}
\caption{Cohesive to partially fractionalised critical behaviour in the broken phase. The dashed black line indicates the edge of the superconducting dome and the coloured lines are some slices at fixed $\Phi_1/\mu$ near the critical solution (black). 
The black dot shows the critical $T=0$ solution with good agreement with the critical $T>0$ solution. Note that the fixed $\Phi_1$ slices in the bottom left panel are slightly curved, signifying a small variation of the charge with temperature at fixed chemical potential and fixed UV deformation $\Phi_1$.
\label{fig:critplot}}
\end{center}
\end{figure}
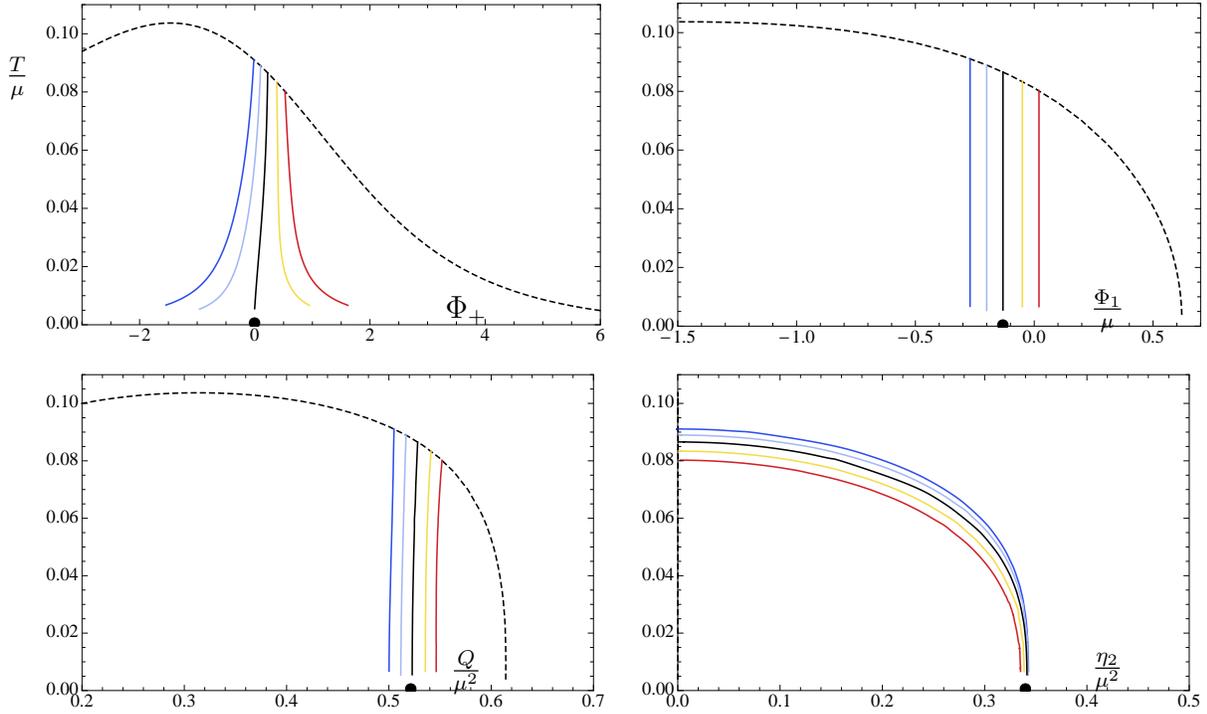
We can illustrate the cohesive to partially fractionalised critical behaviour at finite $T$ by approaching $T=0$ from the top of the superconducting dome (represented by a black dashed line in each case) at constant values of $\Phi_1/\mu$. A selection of such slices is shown in Fig. \ref{fig:critplot}. We see that the critical solution approaches the unique constant dilaton solution in the IR, while solutions on either side of it exhibit an IR divergence as $T=0$ is approached, consistent with our earlier analysis.

\begin{figure}[h!]
\begin{center}
\includegraphics[width=0.75\textwidth]{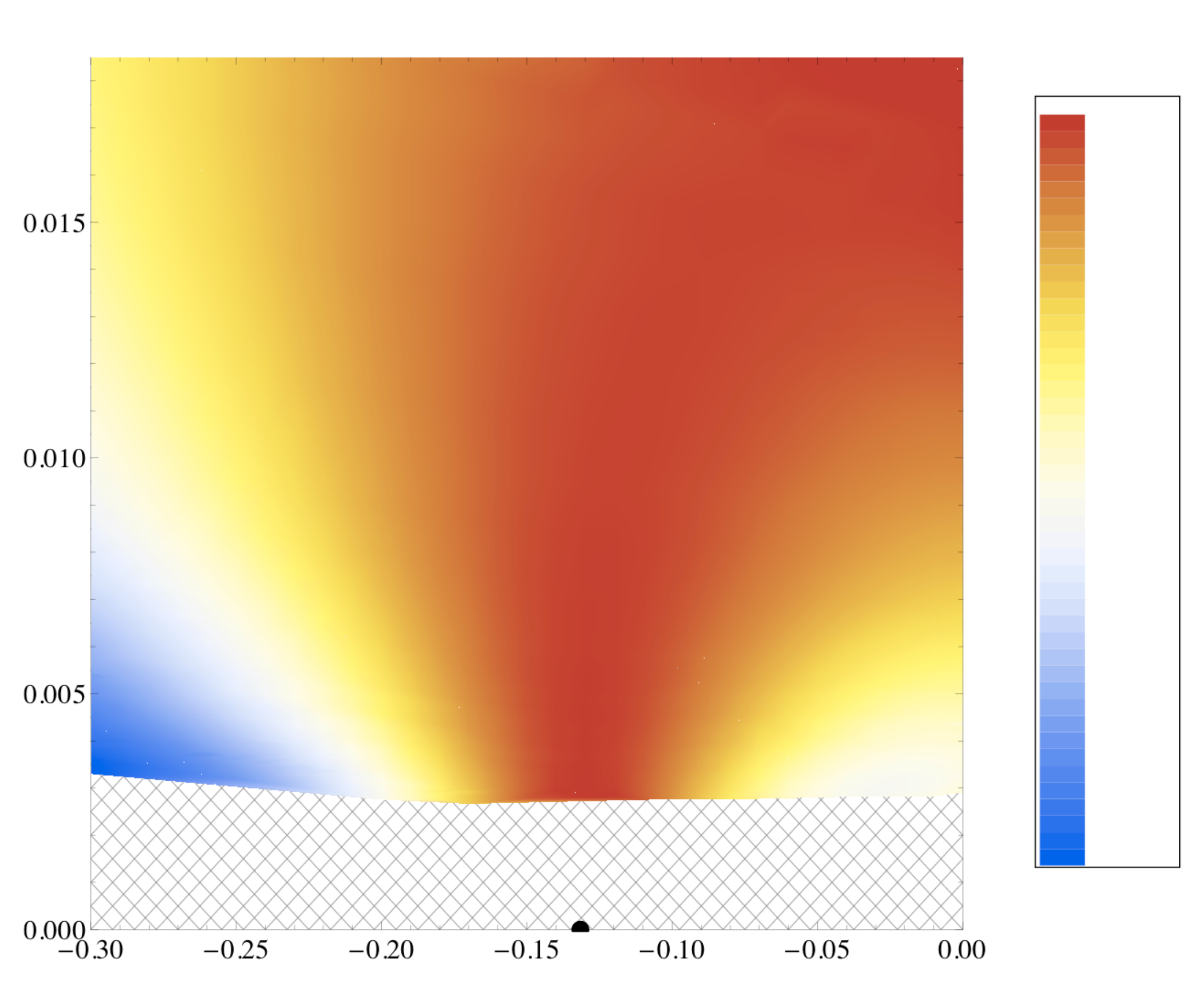}
\setlength{\unitlength}{0.1\columnwidth}
\begin{picture}(0.1,0.25)(0,0)
\put(-7.8,4.){\makebox(0,0){$\frac{T}{\mu}$}}
\put(-3,0.1){\makebox(0,0){$\frac{\Phi_{1}}{\mu}$}}
\put(-0.63,5.3){\makebox(0,0){$1.0$}}
\put(-0.63,0.9){\makebox(0,0){$0.75$}}
\put(-3.7,0.4){\makebox(0,0){QCP}}
\end{picture}
\caption{Scaling behaviour for the quantum-critical region associated with the $z=1, \theta=0$ QCP. Colour indicates the value of $\alpha$, which is proportional to the logarithmic derivative of entropy with respect to temperature at fixed $\mu$. The $T=0$ QCP is located at the  black dot. Hatching indicates low temperature regions not covered by our data. \label{fig:a1density}}
\end{center}
\end{figure}
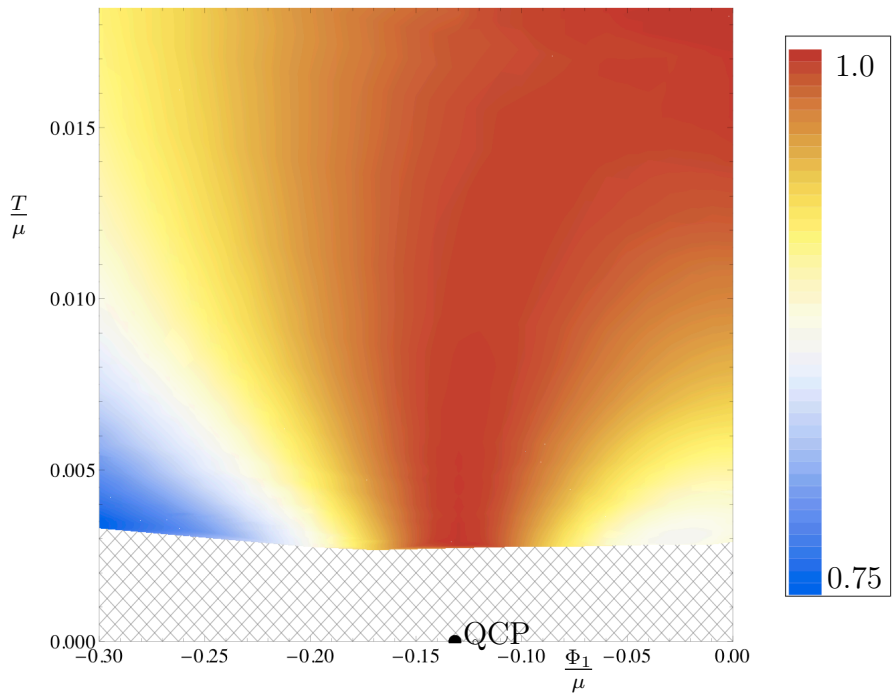

A comprehensive visualisation of the critical behaviour is given in Fig. \ref{fig:a1density} where we show a fit to the scaling $\alpha$ of the entropy density of the form
\be\label{eq:entropyscale}
	\frac{s}{\mu^2} \,\sim\, c_1\left(\frac{T}{\mu}\right)^{\frac{2}{\alpha}} \,,
\ee
for the region below the superconducting dome. From equation \ref{eq:entropy scaling} we expect that near $T=0$:
\be
	\alpha \,\sim\, \frac{z}{1-\frac{\theta}{2}}\,.
\ee
The emergence of the quantum critical wedge associated with the $z=1$ quantum critical point of the cohesive to partially fractionalised ground state is clearly visible. On either side of the wedge, the values of the coefficient $\alpha$ are in agreement with the predicted values of $\alpha=2/3$ in the cohesive phase and $\alpha=2$ in the (partially) fractionalised phases.

\subsection{The Fractionalisation Transition}

Our analysis thus far can be taken as strong evidence for the existence of a $T=0$ phase transition between a superfluid cohesive phase and a superfluid fractionalised phase at a critical UV deformation $\Phi_{1}/\mu(g_m)\simeq-0.131$. In this section we complete the physical picture, deforming the IR geometries of section \ref{sec:T=0} by irrelevant deformations in order to construct full solutions which asymptote to AdS$_4$ in the UV. 

Our numerical results are conclusive: we find one-parameter families of full geometries of each of the three types --- superfluid cohesive phase, superfluid fractionalised phase and fully fractionalised. The superfluid cohesive phase exists for $\Phi_{1}/\mu<\Phi_{1}/\mu(g_m)$. The superfluid fractionalised phase exists for $\Phi_{1}/\mu(g_m)<\Phi_{1}/\mu<\Phi_{1}/\mu(g_f)$, meeting the fully fractionalised branch smoothly at $\Phi_{1}/\mu(g_f)\simeq 0.621$.

\begin{figure}[h!]
\begin{center}
\includegraphics[width=0.48\textwidth,keepaspectratio]{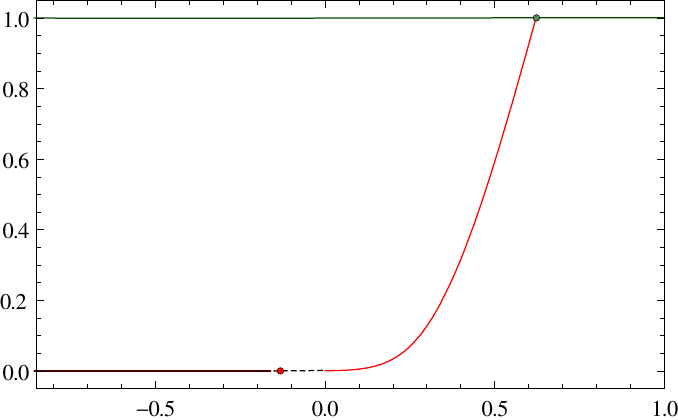}\,
\includegraphics[width=0.48\textwidth,keepaspectratio]{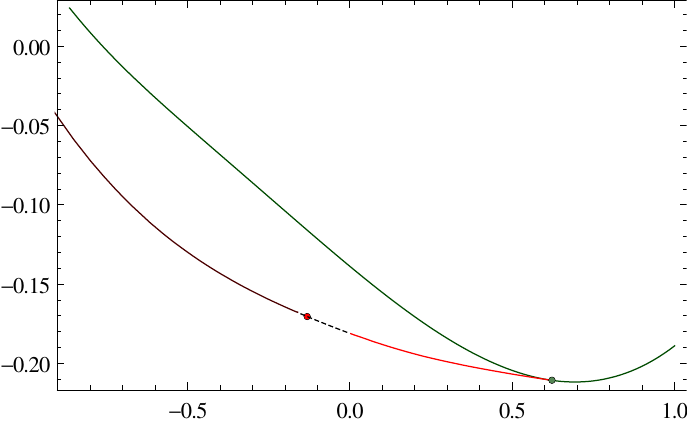}
\setlength{\unitlength}{0.1\columnwidth}
\begin{picture}(0.1,0.25)(0,0)
\put(-10.1,1.75){\makebox(0,0){$\tfrac{\cal{A}}{Q}$}}
\put(-4.95,1.75){\makebox(0,0){$\tfrac{\omega}{\mu^3}$}}
\put(-1.9,0.0){\makebox(0,0){$\Phi_{1}/\mu$}}
\put(-6.9,0.0){\makebox(0,0){$\Phi_{1}/\mu$}}
\put(-2.8,0.55){\makebox(0,0){$g_c$}}
\put(-1.05,0.45){\makebox(0,0){$g_f$}}
\put(-7.9,0.5){\makebox(0,0){$g_c$}}
\put(-5.95,2.65){\makebox(0,0){$g_f$}}
\end{picture}
\caption{Dependence of the $T=0$ domain walls on the parameter $\Phi_{1}/\mu$. As in Fig \ref{fig:summary}, shades of red indicate broken $U(1)$ symmetry, with dark red indicating a superfluid cohesive and bright red a  superfluid fractionalised phase. Fully fractionalised geometries are shown in green. Supplementary low temperature data ($T\simeq 10^{-3}\mu$) in the vicinity of the fractionalisation transition $g_c$ is indicated by black dashes.\label{fig:T=0}}
\end{center}
\end{figure}
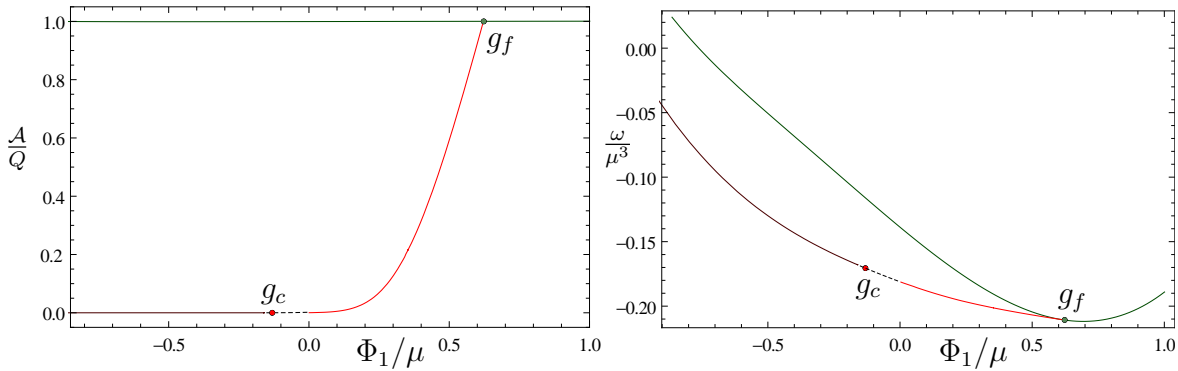

A holographic measure of fractionalisation is the ratio of flux emanating from the deep IR of the solution ({\it e.g.} a black-hole horizon) and the total charge Q:
\be
\frac{\cal A}{Q} \,\equiv\, \frac{1}{Q}\left(\sqrt{-g} Z(\Phi) F^{rt}\Bigr|_{r_+}\right)\,.\label{eq:fracCharge}
\ee
As a consequence of Gauss's Law, this must interpolate between zero for the superfluid cohesive phase and unity in the fully fractionalised phase \cite{Hartnoll:2011pp,Iqbal:2011bf,Huijse:2011ef}. In Fig. \ref{fig:T=0} we plot this measure for each of the solution branches found, which confirms their identities as either superfluid cohesive,  superfluid fractionalised or fully fractionalised.

For the above phase transitions to occur, it is necessary that the fully fractionalised branch be thermodynamically subdominant wherever it coexists with the superfluid cohesive or superfluid fractionalised branches. To this end, we plot the free energy density $\omega$ of each family in Fig. \ref{fig:T=0}. We see that the free energy is indeed higher in the fully fractionalised phase. It proved numerically unprofitable to calculate the $T=0$ free energies of the cohesive and  superfluid fractionalised branches in a small region close to the critical point, so in that region we have supplemented our results with low temperature ($T\simeq 10^{-3}\mu$) data. Furthermore we note the precise agreement between the $T=0$ analysis and the low temperature results, which is thanks to the plateauing of the thermodynamic quantities near $T=0$ that we (numerically) observed in Fig. \ref{fig:critplot}.

The behaviour of the free energy in Fig. \ref{fig:T=0} strongly suggests that the phase transition at $g_c$ is continuous. Indeed, we have computed the first two derivatives of the free energy with respect to $\phi_1$ and have found them to be continuous about $g_c$, implying at least a third-order transition. Our supplementary data does not allow us to study arbitrarily high derivatives with sufficient accuracy.

In contrast, we find a kink in the first derivative of the free energy at $g_f$, implying a second-order transition there. This tallies well with the fact that, in our model, the transition at $g_f$ is one between a superconducting phase and a normal phase. 

Both these results can be compared to the case of fermionic fractionalisation transitions \cite{Hartnoll:2011pp}, where the roles are reversed. The analogous transition between their `mesonic' (here called `cohesive') and partially fractionalised phases are first or second order, depending on the parameters of the model, while the transition between partial and full fractionalisation is continuous of third order.


\subsection{Class \textit{Ib}: A Bottom-Up Model}
The results we have presented so far map out the detailed phenomenology for a theory in class \textit{Ia}, under the classification of Fig. \ref{fig:summary}, for which the coupling term $Z_F$ remains finite for an interval of UV scalar deformations $\Phi_1$. Consequently, we argue, the theory in this interval was able to support superfluid cohesive solutions which provided the thermodynamically preferred solutions.

Here we make a few brief comments on a model governed by the same potential \eqref{eq:ZV}, but with a $\Phi$-even gauge coupling term $Z_F(\Phi) = Z_0^2 \cosh\left(\frac{a\Phi}{\sqrt{3}}\right)$ with $a>0$. The important feature of this model is that $Z_F$ will diverge in the IR given an IR divergent $\Phi$, irrespective of its sign. Thus, the interval of cohesive solutions exhibited by our earlier model will here be reduced to a single point, $\Phi_1=0$. At this point the bulk solution has $\Phi(r)=0$ everywhere and is simply an AdS$_4$ to AdS$_4$ charged domain wall.

Thus our model exhibits a transition from superfluid cohesive phase at the point $\Phi_1=0$, to superfluid fractionalised phase $0<|\Phi_1|<\Phi_1(g_f)$, and ultimately to a fully fractionalised phase when $|\Phi_1|\geq\Phi_1(g_f)$. This is illustrated in Fig \ref{fig:summary}. We now turn to a model which arises from a consistent truncation of 11D supergravity, which shares even gauge coupling $Z$ with class $Ib$, but which differs from all models in class I by having finite $Z$ everywhere. As we shall see this leads to markedly different phenomenology.

\section{M-Theory}
\label{sec:m-theory}

We work with the consistent truncation of \cite{Gauntlett:2009bh,Gauntlett:2009zw}, whose equations of motion can be obtained from the action
\bea\label{eq:MtheoryLagrangian}
S_M &=& \frac{1}{16 \pi G}\int d^4 x\sqrt{-g} \Bigl[R - \frac{(1-h^2)^{3/2}}{1 + 3 h^2}F_{MN}F^{MN} - \frac{3}{2(1-\frac{3}{4}|\chi|^2)^2}|D\chi|^2 \Bigr. \nn
&&\Bigl. -\frac{3}{2(1-h^2)^2} (\nabla h)^2 - \frac{6}{\ell^2}\frac{(-1 + h^2 + |\chi|^2)}{(1 - \tfrac{3}{4} |\chi|^2)^2(1-h^2)^{3/2}}\Bigr] + \frac{1}{16\pi G}\int \frac{2h(3+h^2)}{1+3h^2}F\wedge F\,.\nn
\eea
By defining $\chi=\xi e^{iq\varphi}$ and performing the transformation
\be\label{eq:MtheoryRepara}
h = \tanh\left(\frac{\Phi}{\sqrt{3}}\right)\,,\qquad \rho = \frac{2}{\sqrt{3}}\tanh\left(\frac{\eta}{\sqrt{2}}\right)\,,
\ee
this action falls into the class II of models of (\ref{eq:GenLag}, \ref{eq:complexScalarTransform}) with the specific choices
\bea
Z_F(\Phi) = \frac{4}{\cosh^3\left(\frac{\Phi}{\sqrt{3}}\right)\left(1+3\tanh^2\left(\frac{\Phi}{\sqrt{3}}\right)\right)}\,,
\qquad X(\eta) = \frac{1}{\sqrt{2}}\sinh\left(\sqrt{2}\eta\right)\,
\eea
and charge $q\ell = 1$, \cite{Gauntlett:2009dn}.
This model also has a Chern-Simons term. We now turn to a description of the detailed phase structure of the CFT duals of this system, held at finite density, with particular regard to the zero-temperature ground states. Much of this discussion has been reported in the literature \cite{Gauntlett:2009dn,Gauntlett:2009bh}, and more details can be found there.

\subsection{Phases}
\begin{figure}[th]
\begin{center}
\includegraphics[width=0.6\textwidth]{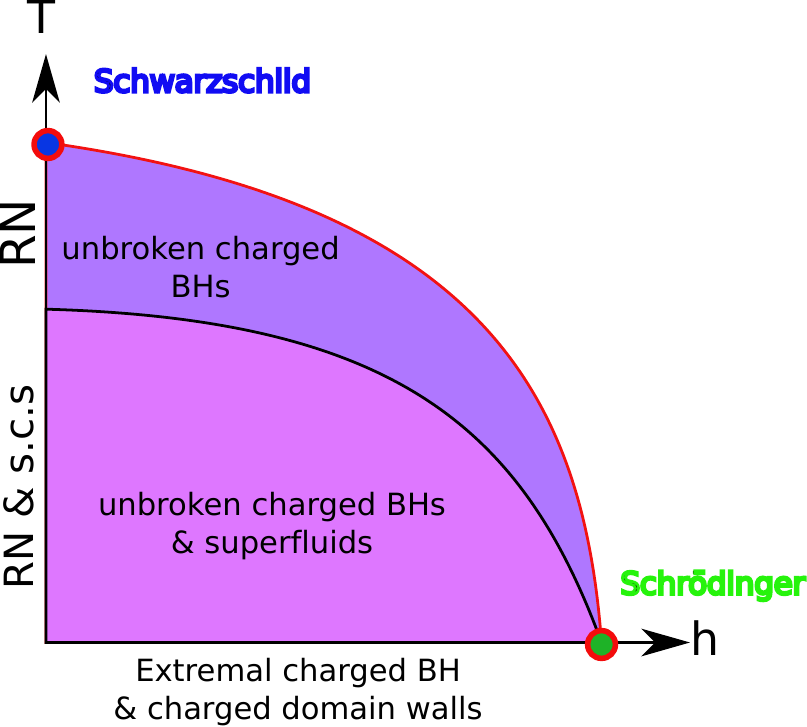}
\caption{The existence of bulk solutions at fixed mass. The red line denotes neutral solutions terminating in the $Schr$ zero-termperature fixed point. Note that the neutral dome meets the superconducting dome at precisely this fixed point, showing that there is no fractionalised phase in this model.
\label{fig:phasediagConstEn}}
\end{center}\end{figure}

As before, a superfluid branch of solutions emerges as an instability of the charged Reissner-Nordstr\"om family of solutions which exists in this model when $h=0$. Once again, under the dilaton deformation in the constant $\mu$ phase diagram we see a dome of superconducting solutions, whose $T=0$ limit, as shown in \cite{Gauntlett:2009dn,Gauntlett:2009bh}, is a charged domain wall interpolating between two AdS regions of different AdS lengths. Whenever the $U(1)$ symmetry remains unbroken the zero-temperature limit of the neutral and charged black hole solutions is incompressible, in the sense that it does not depend on the value of $\mu$. In fact it approaches a zero-density state with vanishing (direct) conductivity. Furthermore, the $T\rightarrow 0$ limit of the neutral and charged solutions meet at the unique, unbroken $T=0$ solution of the system, which lifts to an eleven-dimensional Schr\"odinger solution of M-theory \cite{Kim:2011fb}. Thus the entire region\footnote{The possible emergence of non-isotropic phases \cite{Donos:2011bh, Donos:2011ff} is beyond the scope of this work, but could modify the picture in some regions.} at $T=0$ outside the dome is degenerate on a constant-$\mu$ phase diagram. In view of this `incompressibility' of the groundstate, Fig. \ref{fig:phasediagConstEn} shows the domains where these solutions exist at fixed energy rather than fixed chemical potential as in \cite{Gauntlett:2009bh}.

As we tune the deformation by ${\cal O}_h$, both at $T=0$ and at $T\neq 0$, the expectation from our previous results in this paper is that we eventually reach a transition from the broken $U(1)$ solution to a (partially) fractionalised phase. 

 We now turn to an exploration of the neutral solutions of \reef{eq:MtheoryLagrangian} outside the dome in which the neutral scalar $h$ attains the singular value $h=1$ in the IR. In the parametrisation of Eq. \reef{eq:MtheoryRepara} we see that this corresponds to a (logarithmically) divergent dilaton singularity, which should by now be no surprise.
\subsection{Neutral top-down solutions}%
We rewrite \reef{eq:MtheoryLagrangian} for the case of interest assuming the charged scalar $\chi$ is trivial, since we do not want solutions in which the $U(1)$ symmetry is broken. Hence, passing to the variables \reef{eq:MtheoryRepara}, we find
\be\label{eq:NeutralM}
S = \int d^4x \sqrt{-g} \Bigl[ R -  \frac{{\rm sech^3}\left( \frac{\Phi}{\sqrt{3}} \right)}{1 + 3 \tanh^2 \left( \frac{\Phi}{\sqrt{3}} \right)}F^2 - \frac{1}{2}(\nabla \Phi)^2 +\frac{6}{\ell^2} \cosh\left( \tfrac{\Phi}{\sqrt{3}} \right)  \Bigr]\,.
\ee
Recall that at finite temperature we have a one-parameter family of neutral `dilatonic' black holes, with IR behaviour \reef{eq:FiniteTIR}, where $\xi\equiv 0$. At zero temperature the nature of the solution changes, as we now describe.
Guided by our findings above we suppose\footnote{In support of this assumption we have strong numerical evidence for this behaviour by studying the zero-temperature approach of the none-extremal black hole solutions.} that the isometries of the metric are enhanced from the $\mathbb{R}\times SO(2)$ symmetry above to the full $SO(1,2)$ Lorentz symmetry. By choosing a convenient radial gauge we take
\be\label{eq:IRmetric}
ds^2 = \frac{d\rho^2}{F(\rho)} + \rho^2 {\rm sech}\left(\Phi(\rho)/\sqrt{3}\right) \eta_{\mu\nu}dx^\mu dx^\nu\,.
\ee
This expression is convenient for the current purposes, but we note that we can convert it into the form \reef{eq:GeneralMetricChoice} via
\be
r^2 = \rho^2 \,{\rm sech}\left(\Phi/\sqrt{3}\right)\,,\qquad e^{-\beta} f = r^2\,,\qquad \sqrt{F} = \frac{d\rho}{dr}\sqrt{f}\,.
\ee\label{eq:rhogauge}
We can use this ansatz to construct the singular IR solutions admitted by the M-theory truncation.
\subsubsection{Singular IR solutions}\label{sec:singularIR}
We can construct the desired IR solution by a standard dimensional reduction trick. Suppose then that $\Phi$ diverges\footnote{The case $\Phi\rightarrow - \infty$ is equivalent by the $\Phi \rightarrow -\Phi$ symmetry of the lagrangian coming from $11$D.} in the IR as $\Phi\rightarrow \infty$. In this limit the action approaches
\be\label{eq:lagsing}
S_{\rm sing} = \int d^4 x \sqrt{-g} \Bigl[ R  - 2 e^{-\sqrt{3}\Phi} F^2 - \frac{1}{2}(\nabla\Phi)^2 + \frac{3}{\ell^2} e^{\Phi/\sqrt{3}}\Bigr]\,,
\ee
which can be obtained in a reduction via a gravi-photon ansatz
\be
d\hat s^2 = e^{\Phi/\sqrt{3}} ds^2 (M) + e^{-2\Phi/\sqrt{3}} \left( dz + 2 \sqrt{2} A \right)^2\,,
\ee
where $A\in T^*M$ and hats denote five-dimensional quantities. The starting five-dimensional action is simply Einstein-Hilbert with a cosmological constant:
\be
S_{5} = \int d^5 x \sqrt{-\hat g} \Bigl[ \hat R + \frac{3}{\ell^2} \Bigr]\,.
\ee
Under this reduction, pure five-dimensional AdS with metric
\be
d\hat s_5^2 = \ell_5^2 \left[ \frac{d\rho^2}{\rho^2} + \rho^2 \left( -dt^2 + d\mathbf{x}^2 \right) + \rho^2dz^2\right]\,,\qquad \left(\ell_5^2=4 \ell^2\right)\,,
\ee
reduces to a logarithmic dilaton solution in four dimensions, that is:
\be\label{eq:ads_reduced}
ds^2 = \ell_5^2 \left[ \frac{d\rho^2}{\rho} + \rho^3( -dt^2 + d\mathbf{x}^2 )\right]\,,\quad {\rm with} \quad \Phi = -\sqrt{3}\log (\rho)\,.
\ee
This solution, as we shall now see, plays the role of the singular IR we are after. Note that in this case the gravi-photon field $A$ is trivial, so that the IR solution is obtained via a simple circle reduction from five dimensions.

Note that we also have a five-dimensional Schr\"odinger metric \cite{Son:2008ye,Balasubramanian:2008dm,Adams:2008wt}, obtained as a simple deformation of the above, by adding a term proportional to the light-cone coordinate $(dx^+)^2$:
\be
d\hat s^2 = \ell_5^2 \left[-\beta^2\rho^4 (dx^+)^2 + \frac{d\rho^2}{\rho^2} + \rho^2 \left( dx_2^2 + 2 dx^+ dx^- \right)\right]\,,
\ee
where $\beta$ is a constant. Once reduced to four dimensions and seen as an expansion in small $\rho$, the Schr\"odinger and AdS metrics in fact {\it agree} to leading order, that is they give the same IR behaviour for the desired logarithmic dilaton solution in four dimensions. Of course, solving the full equations coming from \reef{eq:MtheoryLagrangian} order by order will determine the exact solution and thus distinguish between the two cases. This was pointed out earlier in \cite{Kim:2011fb}. 

Here we demonstrate that this solution is in fact the universal $T\rightarrow 0$ limit of the model \reef{eq:MtheoryLagrangian} outside the dome. The full M-theory solution, lifted to eleven dimensions, is a Schr\"odinger solution ${\rm Schr}_5 \times KE_6$ with dynamical critical exponent $z=2$, where the ${\rm Schr}_5$ is fibred\footnote{Similar solutions were constructed previously in \cite{O'Colgain:2009yd}.} non-trivially over the K\"ahler-Einstein space $KE_6$. In the dual field theory this solution corresponds to a mass-deformation driving the system to the non-relativistic $z=2$ IR fixed point \cite{Kim:2011fb}. The full structure of this is not evident from the four-dimensional perspective, but we see the $z=2$ scaling for example in the temperature scaling of entropy density at criticality.

Returning thus to constructing the neutral IR solutions, we substitute the metric ansatz into the equations following from \reef{eq:MtheoryLagrangian}, resulting in a single, fully decoupled, non-linear ODE for the scalar field $\Phi$. Its precise form is given in appendix \ref{sec:appODE}. Once $\Phi$ is determined, the Einstein equations determine the metric function $F(\rho)$ algebraically:
\be
F(\rho) = \frac{4 \ell^{-2} \rho^2 \cosh\left(\Phi/\sqrt{3}\right)}{4 - (\rho\Phi'/\sqrt{3})^2{\rm sech}^2\left(\Phi/\sqrt{3}\right) - 4 \rho \Phi'/\sqrt{3}\tanh\left(\Phi/\sqrt{3}\right)}\,,
\ee
where the prime denotes a derivative with respect to the radial direction $\rho$. The $\Phi$ equation admits an IR expansion, which reproduces \reef{eq:ads_reduced} to leading order. It takes the form
\be
\Phi(\rho) = -\sqrt{3}\log \rho + \frac{3\sqrt{3}}{2}\rho^2 + \cdots\,,
\ee
and contains no free parameters. Note that the freedom associated with the $\rho$ rescaling of the metric has no physical effect on the solution. This unique IR expansion can be integrated to the UV and connects to the desired skew-whiffed $AdS_4$ solution. We have done this and read off the UV data. In appropriate units\footnote{We quote numerical values in the same convention as \cite{Gauntlett:2009bh} for ease of comparison.} one finds
$h_2/h_1^2 =- 0.288\ldots \sim-1/(2\sqrt{3})$ for the expectation value $\langle {\cal O}_h\rangle$ and $\varepsilon/h_1^3 = 0.779\ldots \sim\frac{3}{3\sqrt{3}}$ for the energy. The analytic values are those of the exact solution found in \cite{Gauntlett:2009bh}. One can check that the chemical potential $\mu(r_\infty) - \mu(r_{\rm IR})$ vanishes. We thus conclude that the unique neutral $T=0$ solution of the theory \reef{eq:MtheoryLagrangian} with logarithmically diverging dilaton is the analytic solution given in Eq (8.2) of \cite{Gauntlett:2009bh}.

\subsection{Approaching the critical point}
It is enlightening to study the power-law associated with the approach to zero temperature. At the critical value we see scaling behaviour compatible with the $z=2$ fixed point, in other words $T$-linear scaling.
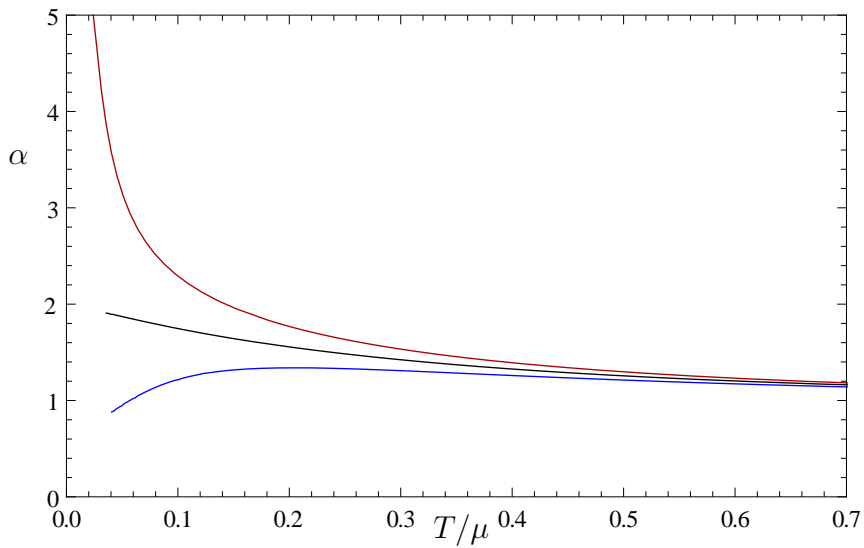
\begin{figure}[th]
\begin{center}
\includegraphics[width=0.7\textwidth]{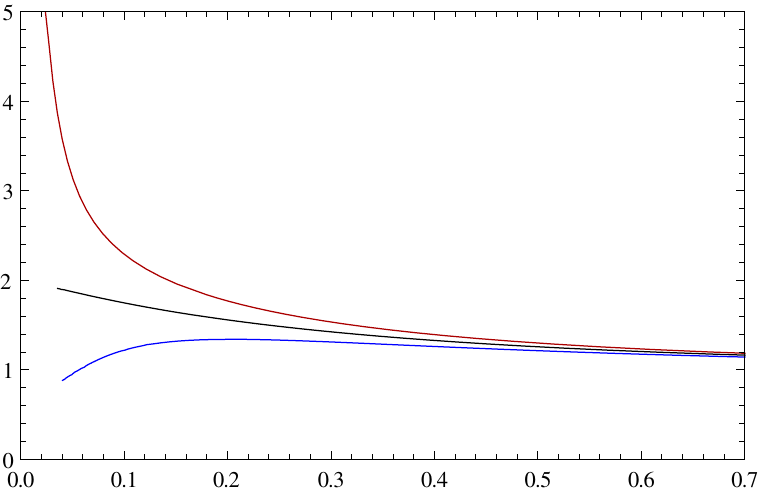}
\setlength{\unitlength}{0.1\columnwidth}
\begin{picture}(0.1,0.25)(0,0)
\put(-3.6,-0.0){\makebox(0,0){$T/\mu$}}
\put(-7.4,3.2){\makebox(0,0){$\alpha$}}
\end{picture}
\caption{Approaching the M-theory Schr\"odinger critical point from finite temperature. The black line for $h_1/\mu =h_{1c}/\mu =0.354...$ shows precisely the critical $T$-linear scaling. It delineates the blue supercritical line ($h_1/\mu = 0.382$) from the red subcritical line ($h_1/\mu = 0.324$). Note that the complex scalar field is set to zero in all cases here, and the subcritical region is masked in the full theory by a superconducting dome.
\label{fig:mcritical}}
\end{center}\end{figure}
Consulting Fig. \ref{fig:mcritical} we see that this scaling behaviour is indeed true precisely, if the temperature is lowered holding $\Phi_1/\mu$ at the critical value. The critical curve in black clearly approaches $\alpha=2$ in the appropriate limit. The red curve corresponds to the approach of the (unstable) $AdS_2$ infrared, and so the diverging $z\rightarrow \infty$ is expected. Note that this region is cloaked by the superconducting dome in the full phase diagram. The blue curve corresponds to a value of $h_1/\mu$ outside the dome. The value of $\alpha$ remains bounded as $T$ approaches zero in this case. While the zero-temperature solution itself, for $h_1/\mu>h_{1c}/\mu$ is independent of the value of the chemical potential the finite-temperature solutions are not, and so the {\it approach} does depend on ratios like $h_1/\mu$ or $T/\mu$. This is clearly illustrated in Fig. \ref{fig:mcritical}, where the critical scaling with $\alpha=2$ delineates a region of supercritical (as exemplified by the blue curve) scaling with $\alpha<2$ from a region of subcritical scaling with $\alpha\rightarrow\infty$, as exemplified by the red curve. The latter behaviour corresponds to approaching the zero-temperature AdS$_2$ geometry, which in the full theory is masked by the superconducting dome.

\section{Discussion}
The results of this paper are twofold. On the one hand we have constructed bosonic analogues of so-called `fractionalisation transitions'  by relying on suitably constructed minimal bottom-up models. On the other hand we reexamined the phase diagram of the M-theory superconductor in this light, showing how the analogous transition in this system is in fact a transition from a superfluid cohesive phase to a neutral phase with vanishing (direct) conductivity\footnote{Interestingly, however, there is a non-vanishing transverse conductivity ${\rm Re}\sigma_{xy}\neq 0$.}. We emphasise again, that these transitions are taking place at constant chemical potential.

The IR behaviour of the dilaton coupling to the gauge kinetic term at zero temperature governs the quantum phase transitions observed in this paper. All examples encountered in this work support the conjecture that one cannot have a cohesive phase if the dilaton coupling to the gauge-kinetic term diverges ($Z\rightarrow\infty$). While we do not have a rigorous argument that would prove this, we can see heuristically that it makes sense: diverging $Z$ means that the effective gauge coupling vanishes as is obvious in a normalisation where the gauge-kinetic terms reads $\sim \frac{1}{e^2}F^2$. That means that matter in the strict IR, where $Z$ diverges, cannot source any flux. Thus any flux there can only emanate from a horizon and consequently we would have a fractionalised contribution to the overall flux. Our arguments for this behaviour are purely holographic, that is we use the dynamics of the bulk gravity. Similarly our classification of the different phases are based on such bulk arguments. Clearly it would be very interesting to use such bulk arguments to shed more light on the nature of a possible order parameter for the fractionalisation transitions of the kind exemplified in this paper.

In asymptotically $AdS$ spacetimes, one often finds that the zero temperature limit of a black hole solution has an emergent $AdS_2$ geometry, indicated by a divergent dynamical critical exponent. The dual field theory then has a non-zero entropy at zero temperature. Often these finite entropy geometries are unstable when embedded in top-down models (see e.g. \cite{Gauntlett:2009dn,Gauntlett:2009bh,Donos:2011bh}), so that the third law of thermodynamics is upheld by the classical geometry.

As is clear from equation \reef{eq:entropy scaling}, and strikingly visualised in Fig. \ref{fig:a1density}, the $T=0$ infared hyperscaling parameter governs the scaling behaviour of various thermodynamic quantities at low temperature. In particular, a divergent critical exponent can be compensated for by an equally divergent hyperscaling parameter \cite{Hartnoll:2012wm}, resulting in a vanishing entropy at $T=0$. In this manner one can avoid instabilities associated with finite entropy. Our bottom-up model of section \ref{sec:bottom-up} uses precisely this mechanism. The M-theory model, however, is of class II, and so there is no fractionalised phase associated with a vanishing gauge coupling and its corresponding hyperscaling violation solutions; we indeed see that the fractionalised (Reisner-Nordstr\"om) phase is unstable to scalar condensation, and so masked by a superconducting dome. It would be very interesting to investigate precisely what conditions have to be met for any entropic singularity (\textit{i.e.} any $AdS_2$ IR geometry) to be unstable \cite{Iqbal:2011aj} to new phases cloaking it at low temperatures.

Recently, following earlier work on Lifshitz geometries \cite{Harrison:2012vy}, it has been suggested \cite{Bhattacharya:2012zu,trivedi2012} that hyperscaling IR geometries are destabilised due to the running dilaton. If this is the case for the geometries considered here then our results will still apply for an intermediate range of energy scales, above the scale where quantum corrections smooth out the IR geometry. It is interesting in this context that at least some of these seemingly singular geometries can be lifted to perfectly regular higher-dimensional geometries (see also \cite{Gouteraux:2011ce}), potentially eliminating the need to consider the corrections of \cite{Bhattacharya:2012zu,trivedi2012}. Note, however, that in the case of Lifshitz geometries, the higher-dimensional lifts were found to be as problematic as their lower-dimensional descendants \cite{Horowitz:2011gh}, as manifested in the propagation of test strings. The status of the possible resolution of such singularities (e.g. by matter sources \cite{Bao:2012yt}) in string theory deserves further attention.

\subsection*{Acknowledgements}
\noindent
It is a pleasure to thank Aristomenis Donos, Jerome Gauntlett, Andrew Green, Sean Hartnoll, David Tong and Subir Sachdev for discussions. JS would like to thank the MCTP at the University of Michigan and the CTP at the Massachusetts Institute of Technology for hospitality while this work was in progress. AA and BC are supported by STFC studentships. BW is supported by the Royal Commission for the Exhibition of 1851.
\appendix
\section{The ODE in section \ref{sec:singularIR}\label{sec:appODE}}
Here we give the details of the ODE determining $\Phi(\rho)$ in section \ref{sec:singularIR}, which follows from the lagrangian \reef{eq:lagsing} upon setting $A_t = \mu$ with $\mu$ constant.
\bea
\rho^2\Phi''(\rho) &=& \frac{-1}{2\sqrt{3}}\tanh\left( \tfrac{\Phi(\rho)}{\sqrt{3}}\right) \left[12 - \rho^2\Phi'(\rho)^2 \left( 4 + 3\, {\rm sech^2\left( \tfrac{\Phi(\rho)}{\sqrt{3}} \right)} \right) \right]\nn
 &&- \left[ 1 + 3\, {
\rm sech}^2\left( \tfrac{\Phi(\rho)}{\sqrt{3}}  \right) \right]\rho\,\Phi'(\rho) +\frac{1}{12} {\rm sech}^4\left( \tfrac{\Phi(\rho)}{\sqrt{3}}\right)\rho^3 \Phi'^3(\rho)\,.
\eea
For any solution the chemical potential in the boundary theory is arbitrary, as $A$ is pure gauge in the bulk. If $A$ is to be manifestly well-defined in Kruskal coordinates one should choose the gauge $A_t=0$.
%

\bibliography{FracBib}{}
\bibliographystyle{utphys}

\end{document}